\documentclass[11pt]{article} 
\usepackage{osid,graphics,graphicx,axodraw,subfigure}
\usepackage{lscape,amsfonts,makeidx,bm,eucal,tensor,mathrsfs,color,hyperref,amssymb}

\newcommand{\ket}[1]{| #1 \rangle}
\newcommand{\braket}[2]{\langle #1 \vert #2 \rangle}

\hypersetup{hyperindex,backref,pdfpagemode=FullScreen,colorlinks=flase} 

\usepackage{cite}

\def\ket#1{\mathinner{|{#1}\rangle}} 
\def\braket#1{\mathinner{\langle{#1}\rangle}} 
 
 {\catcode`\|=\active \gdef\Braket#1{
\begingroup \mathcode`\|32768\let|\BraVert\left<{#1}\right>
\endgroup}} 
\def\BraVert{\egroup\,\mid\,\bgroup} 

\title{George Sudarshan and Quantum Dynamics}
\author{Kavan Modi
\\{\footnotesize\it School of Physics and Astronomy, Monash University, Victoria 3800, Australia\\ \sc{Email:} kavan.modi@monash.edu}}

\begin{document}
\maketitle

\begin{abstract}
These are my recollections of working with George Sudarshan from 2002 to 2008 when I was a PhD student in his group. During these years I learnt a lot of physics and also witness to some remarkable occurrences.
\end{abstract}
 
\date{\today}


When I began my PhD studies at the University of Texas at Austin in late 2001, I already knew the name George Sudarshan. He was famous for his discovery (as a grad student) of one of the fundamental forces of nature, the weak force. However, before coming to Austin I had spent the summer of 2000 at Fermilab. There I learned at least one thing; I didn't want to study particle physics. Thus I didn't bother looking Sudarshan up. I was interested in complexity and computation and wasn't really interested in the condensed matter group or the nonlinear group. Of course, at the time I didn't know that Sudarshan hadn't worked on particle physics for a long time. And I certainly didn't know of the many other things that he had done, at least two which were worthy of the Nobel Prize.

Ultimately, it was clear that Sudarshan was one of the few theorists interested in the structure of quantum mechanics and the physical world. Working with him was easy in many ways. During the semester we spent either two or three days hanging out in his office for several hours and discussed all sorts of things. During summers, he was usually away, this was a good time for the students to get the writing done. All is all, in the years I spent in his group I got to work on two of these topics and was exposed to, at least to the fascinating history, of a third topic. 

Sudarshan's major achievements include the V--A theory (weak force), inventing the theory of quantum optics, discovering quantum maps, the quantum semi-group master equation, quantum Zeno effect, and tachyons. Yet, many physicists I meet don't know the name George Sudarshan. Some know him for his work on quantum optics, some for open dynamics, but very few know his contributions to particle physics, symmetries, and tachyons. Many will know of the Zeno effect, yet not Sudarshan. In many cases, someone else got the credit. However, it seems to me that in each case for a different reason. I'll give my two cents where I can, and along the way weave in some real science that I've worked on because of my interactions with Sudarshan.

\section{Quantum Zeno and anti-Zeno Dynamics}

My first project under Sudarshan began in late 2002 or early 2003. He had me studying parts of Dirac's book on particle decay. Although I went through a dozen books to understand how researchers have justified exponential decay of particles. Pretty soon it became evident that Sudarshan wanted me to understand the recent experiment by Mark Raizen's group, published just a year ago. This experiment reported the first observation of the \emph{quantum Zeno effect}~\cite{PhysRevLett.87.040402}. 

Quantum Zeno effect was first predicted by Misra and Sudarshan~\cite{bib1, PhysRevD.16.520} back in the late 1970s. It argues that the time evolution of a quantum state, when frequently measured, is hindered. In the limit of continuous measurement the time evolution of the state, in principle, completely stops. For Sudarshan, exponential decay of particles was in contradiction with the rules of quantum mechanics.

Some years prior to the Raizen experiment, direct experimental observation of the quantum Zeno effect was obtained by Itano et al.~\cite{PhysRevA.41.2295}. However, in this experiment the quantum system was not unstable, rather it was an oscillating three-level system. In contrast, the system in the Raizen experiment was really unstable, when left unmeasured its decay was exponential. Sudarshan was certainly happy to see his theory validated nearly 25 years after its inception. (Actually, experimental evidence supporting the quantum Zeno effect in particle physics experiments was first pointed out by one of Sudarshan's PhD student back in late 1970s~\cite{PhysRevD.21.1304, Valanju:1980mi}.) But the Raizen experiment exercised some serious control.

There was a flip side to the matter. In those years several authors had challenged the foundational importance of the Zeno effect. Moreover, they had suggested that the opposite of the quantum Zeno effect may also be true~\cite{PhysRevA.56.1131, Nature.405.546, PhysRevLett.86.2699}. That is, frequent measurements can be used to accelerate the decay of an unstable state, this effect was named as the \emph{anti-Zeno effect} or the inverse Zeno effect. The argument was that the original formulation of the quantum Zeno effect treated the measurement process as an idealised von-Neumann type; that is an instantaneous event that induces discontinuous changes in the measuring system. The anti-Zeno effect was first identified as a possibility when measurement processes that take a finite amount of time were considered.

While the argument for Zeno effect was clean, that for the inverse effect was complicated. Firstly, applying a non-projective measurement destroys the original state for any outcome. Moreover, the complex models for the measurement made things very confusing because one could not separate a technical detail from a fundamental effect. To add to the confusion, the Raizen experiment reported both the Zeno effect and the anti-Zeno effect. Neither side of the argument was happy with this, and Sudarshan wanted someone to look into the matter. The project was led by Anil Shaji, who was the senior student in the group. Anil began supervising me to construct a model to reproduce the results seen in the Raizen experiment. 

The model was simple and based old notes of Sudarshan himself. These notes predated the Zeno effect by nearly two decades. Nevertheless the model reproduced all of the important results of this experiment. It took a bit of fine tuning on my part, but then the results were remarkably clear. However, at the time the situation was very murky. It took us several years to publish these results and there were many blows along the way. Yet, in hindsight, this project was a incredible learning experience.

\subsection{The Raizen Experiment}\label{exp}

In the Raizen experiment~\cite{PhysRevLett.87.040402} the motional states of sodium atoms in a magneto-optical trap, that could be moved, were studied. The atoms were initialised so that they remained inside the moving trap, this stable state was dubbed as the ``ground" state. The state of the atoms was made unstable by accelerating the atoms along with the trap at different rates. When the trap is accelerated the atoms to quantum mechanically tunnel through the barrier into the continuum of available free-particle states. This is the decay process in the experiment.

At the end of the experiment, the spatial distribution of the atoms was recorded to determine the number of atoms that tunnelled out of the bound state inside the trap as a function of time. Because the trap was accelerated throughout the experiment, an atom that spent more time in the and trap would have a higher velocity and thus move farther per unit time. The spatial distribution of the atoms thus contained the information about if and when they tunnelled through the trap.

\begin{figure}[!ht]
\centering
\includegraphics[width=0.48\linewidth]{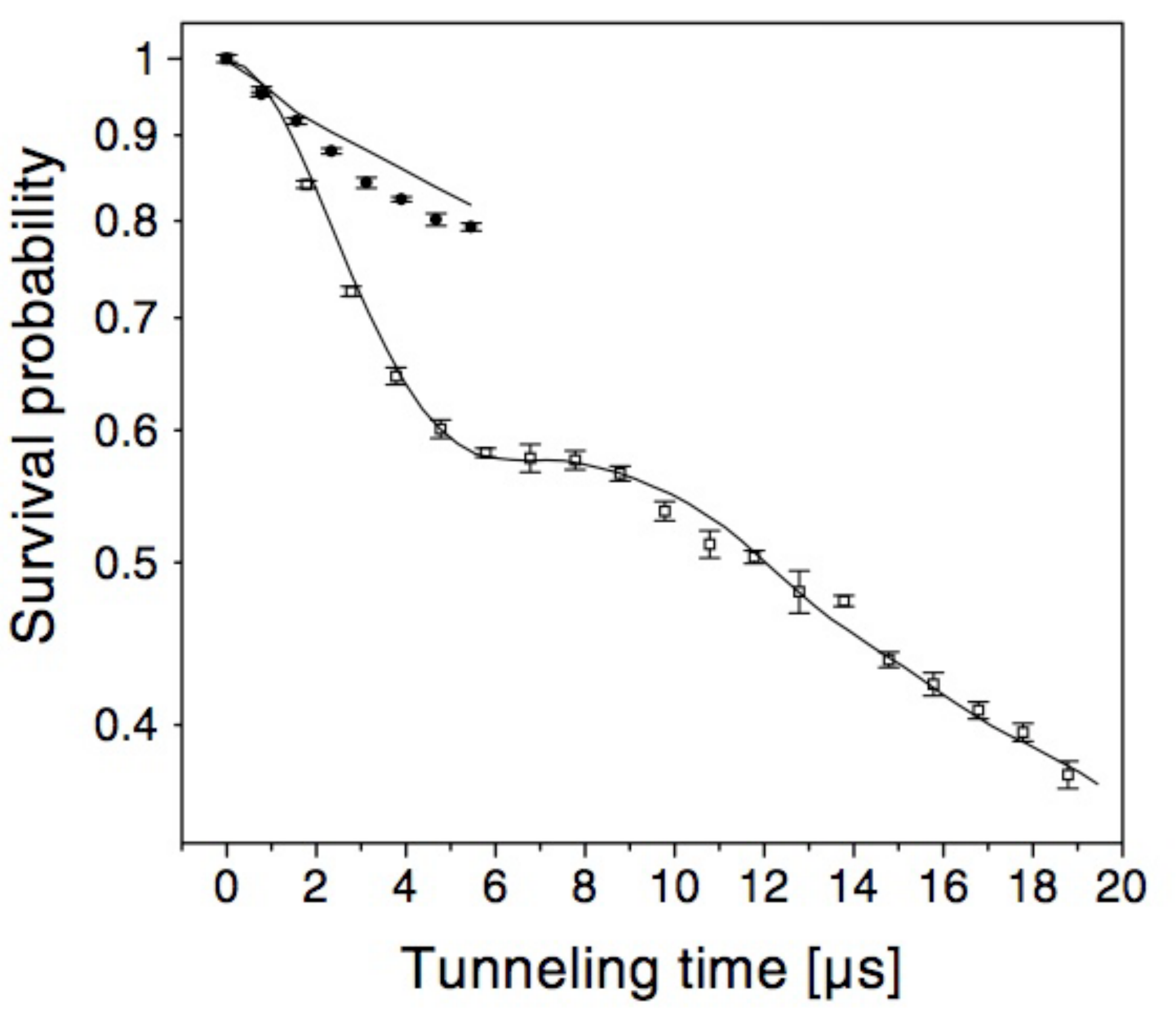}
\includegraphics[width=0.48\linewidth]{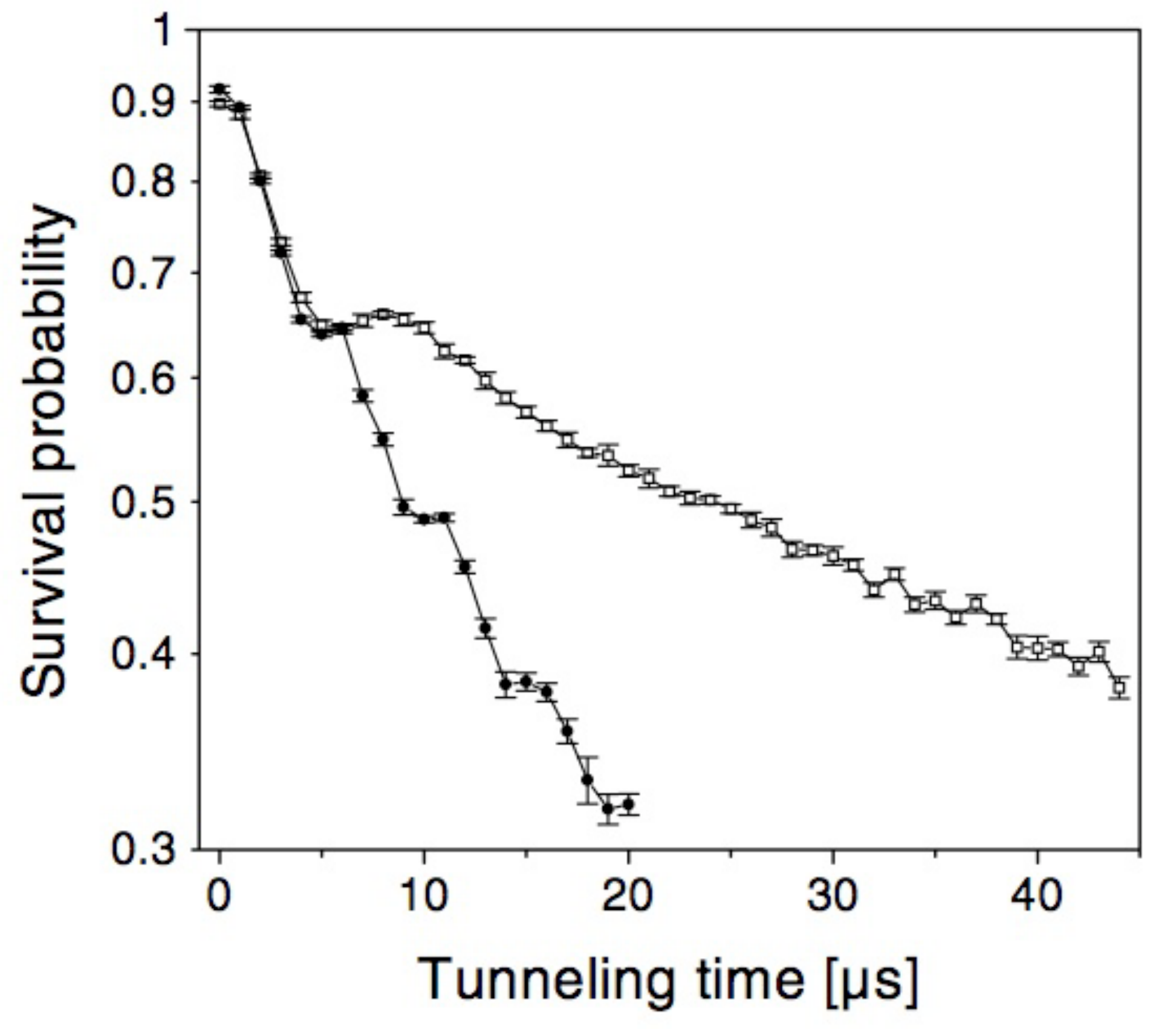}
\caption{The Figures are taken from the paper reporting the Raizen group experiment~\cite{PhysRevLett.87.040402}. The lower line, on the left, is the ``unmeasured" decay curve corresponding to the case where there are no interruptions, and the tunnelling out of the trap is always present. The upper line corresponds to the case where the dynamics are interrupted every 1 $\mu$s leading to the quantum Zeno effect. The upper line, on the right, is the ``unmeasured" decay. The lower line corresponds to the case where interruptions are made every 5 $\mu$s leading to the anti-Zeno effect. \label{sur}}
\end{figure}

The Zeno and the anti-Zeno effects were obtained by interrupting the tunnelling of the atoms out of the trap. The interruption is achieved by lowering the acceleration rate, which in turn lowered the tunnelling rate, for a fixed duration of time. This amounts to a measurement because the tunnelled atoms are separated into groups spatially due to the interruption, i.e., before and after the interruption. Importantly, the interruption periods had to be long enough ($40\mu s$) to spatially resolve the atoms that tunnel out before from the one that tunnelled out after each interruption. Measuring the number of atoms in each group and knowing the total number of atoms that were initially in the trap, one can determine the survival probability of the bound motional states, see Fig.~\ref{sur}.

The Zeno or anti-Zeno effects were obtained by tuning the frequency with which interruption periods were applied. Interruptions once every micro-second led to the quantum Zeno effect, while interruptions rate of $5\mu s$ led to the anti-Zeno effect.

Quantum Zeno effect predicts the zero slope for the survival probability at $t=0$ for a generic unstable quantum state. This feature is seen in the shape of experimentally observed survival probability, i.e., the ``unmeasured system" has an inflection point at $t \ll 1\mu s$ as seen in Fig.~\ref{sur}. Moreover, repeated measurements hinder the decay of the system, this is again as predicted. 

The anti-Zeno effect, on the other hand, is present due to the inflection point at $t\approx 5\mu s$. This is in contrast to the claim that the anti-Zeno effect is more fundamental than the Zeno effect. That is, it only appears at a later time and requires waiting longer to be observed. Yet, it's not clear where this feature comes from. Our paper~\cite{kmzeno} builds a physical model that explains this feature. Namely, we show that if a second unstable bound state is the present then the second inflection point may appear.

This intuition came from Anil Shaji, who had been studying this topic in great details~\cite{shaji04}. (His thesis was not on Zeno dynamics, this was just a sideshow!) While very sensible, we still had to justify this intuition. I started digging around and pretty soon found that in an earlier experiment that led to this experiment, Bharucha et al. observed tunnelling of sodium atoms from an accelerated trap~\cite{PhysRevA.55.R857}. There they say:

\begin{quotation}
\emph{When the standing wave is accelerated, the wave number changes in time and the atoms undergo Bloch oscillations across the first Brillouin zone. As the atoms approach the band gap, they can make Landau-Zener transitions to the next band. Once the atoms are in the second band, they rapidly undergo transitions to the higher bands and are effectively free particles.}
\end{quotation}

It's clear that higher energy states were present in their system. Therefore, an atom in the ground state has to go through the intermediate states before it can tunnel out to the set of free particle states. Now, we present the model that reproduces the experimental observations.

\subsection{A Model for the Raizen Experiment}

We considered an interacting field theory of four fields labelled $A$, $B$, $C$, and $\Theta$. Only $\Theta$ is labelled by a continuous index $\omega$, while other fields are assumed to only have discrete modes. The allowed processes in the model are 
\begin{equation}
A \longleftrightarrow
B \; \mbox{and} \; B \longleftrightarrow
C \ \Theta.
\end{equation}

The Hamiltonian for this model, with these allowed processes, can be written down as,
\begin{equation}
\label{h0} H=H_0+V
\end{equation}
where,
\begin{equation}
\label{h1}
H_0=E_A \ a^{\dagger} a +E_B \ b^{\dagger} b
+\int^{\infty}_{0}d\omega\;\omega\;\theta^{\dagger} (\omega) \theta(\omega) 
\end{equation}
and
\begin{equation}
\label{h2}
V=\Omega \ a^{\dagger} b +\Omega^{*} \ b^{\dagger} a +\int^{\infty}_{0}d\omega 
\left[f(\omega) \ b^{\dagger} c \theta(\omega) + 
f(\omega)^{*} \ c^{\dagger} \theta^{\dagger}(\omega) b \right].
\end{equation}
Here $a^{\dagger}$ ($a$) etc. represent the creation (annihilation) operators corresponding to four fields. The two discrete energy levels are denoted by $E_A$ and $E_B$.

The Hamiltonian in Eq.~(\ref{h0}) is obtained by modifying the Hamiltonian for the Friedrichs-Lee model~\cite{bib11, PhysRev.95.1329}. Sudarshan had studied this model in depth fifteen years before Zeno~\cite{bib10}. Following his notes, the above Hamiltonian can be written in the matrix form as
of $H_{0}$ as basis,
\begin{equation}\label{e1}
H=\left(\matrix{E_A&\Omega^{*}&0\cr
\Omega&E_B&f^{*}(\omega')\cr
0&f(\omega)&\omega \delta (\omega -\omega')\cr}\right).
\end{equation}
Let us represent an eigenstate of $H$ with eigenvalue
$\lambda$ as $\ket{\psi_\lambda}$, satisfying the eigenvalue equation
\begin{equation}\label{eigeq}
H\psi_{\lambda}=\lambda\psi_{\lambda}.
\end{equation}
We express $\psi_\lambda$ also in terms the eigenstates of the bare Hamiltonian $H_0$,
\begin{equation}\label{e2}
\psi_{\lambda}=\left(\matrix{
\braket{A|\psi_\lambda}\cr
\braket{B|\psi_\lambda}\cr
\braket{C\Theta(\omega)|\psi_\lambda}\cr}\right)\equiv
\left(\matrix{\mu^{A}_{\lambda}\cr\mu^{B}_{\lambda}\cr\phi_{\lambda}(\omega)\cr}
\right).
\end{equation}

We are interested in the time evolution of the eigenstates of $H_0$, namely the two bound bare states $\ket{A}$ and $\ket{B}$, and the continuum states $\ket{C\Theta(\omega)}$. The state $\ket{A}$ in our model corresponds to the unstable bound state occupied by the atoms inside the trap in the Raizen experiment~\cite{PhysRevLett.87.040402}. The states $\ket{C\Theta(\omega)}$ represent the continuum outside the trap into which the bound state can decay. 

In other words, the last equation is exactly what we need. Suppose the system is initially prepared in the ground state of $H_0$, i.e., state $\ket{A}$. Once perturbation $V$ (acceleration) is introduced, the first term of the vector represents the survival probability of being in the ground state. Mathematically, the spectrum of the physical Hamiltonian will no longer include bound states. The eigenstates of $H$ belonging to the continuum, corresponding to eigenvalues $0<\lambda<\infty$, will form a complete set of states.

The key difference between Sudarshan's original notes~~\cite{bib10}, and the model we considered is an additional unstable bound state $\ket{B}$ which represents a second bound motional state of the trap. The state $\ket{A}$ is directly coupled to only $\ket{B}$ and the decay of $\ket{A}$ into $\ket{C\Theta}$ is mediated by the new state $\ket{B}$. We showed that the presence of the additional bound state can explain several of the key features of the Raizen experiment~\cite{PhysRevLett.87.040402}.

\subsection{Effective Evolution: Zeno and anti-Zeno}

\begin{figure}[!t]
\centering
\includegraphics[width=0.48\linewidth]{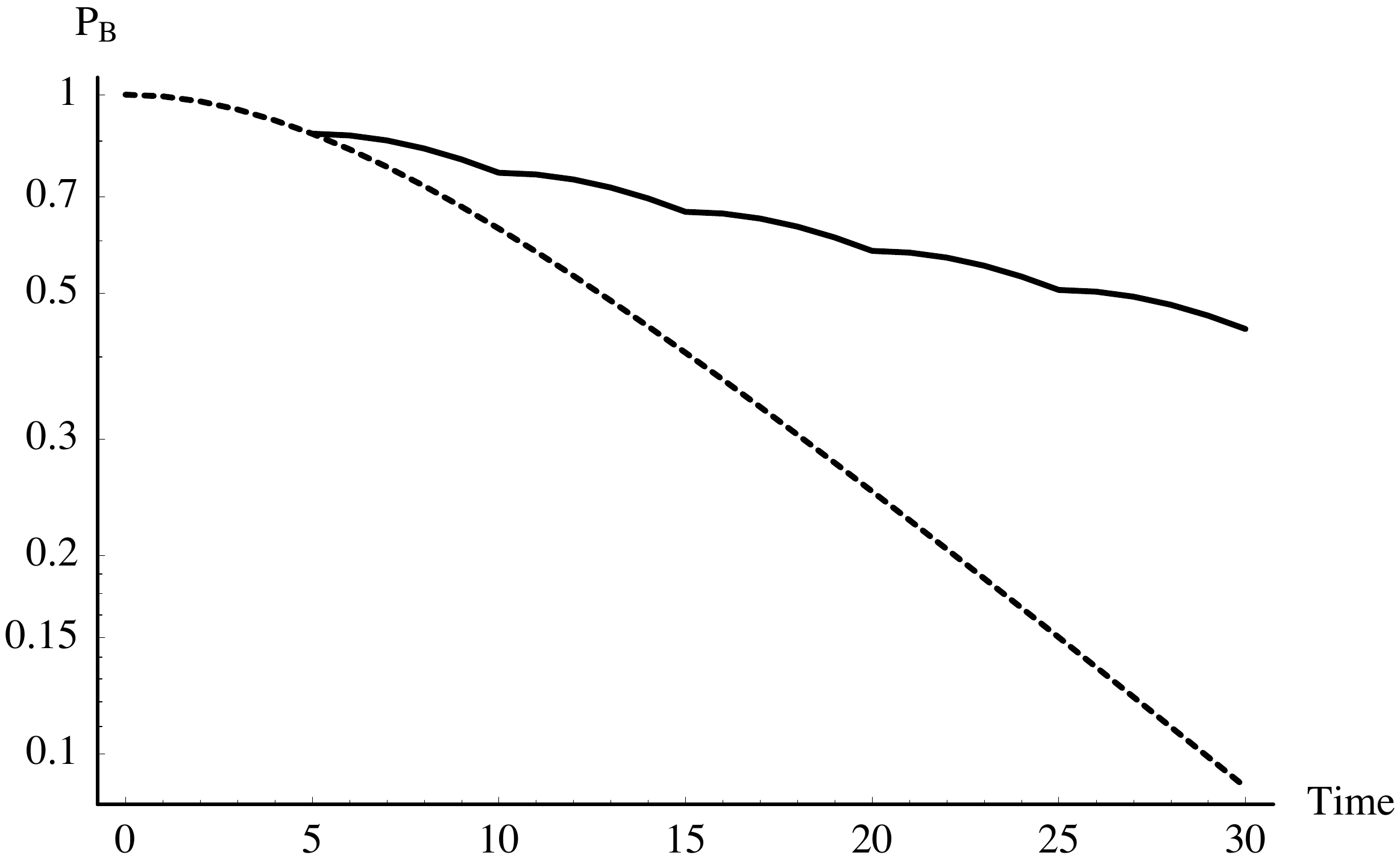}
\includegraphics[width=0.48\linewidth]{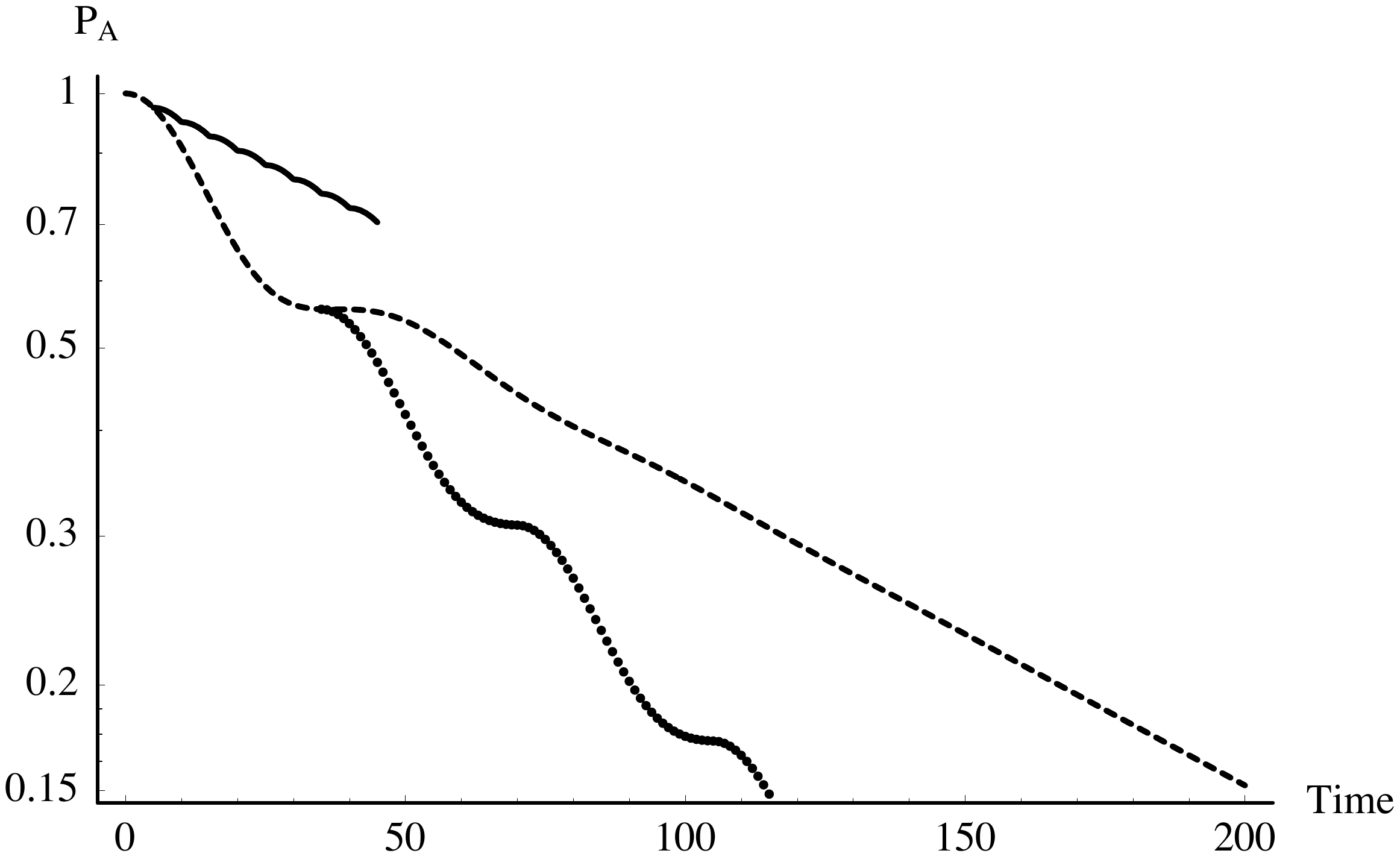}
\caption{(Left) Survival probability of $\ket{B}$. The dashed line shows the ``unmeasured" evolution and the solid line shows the effect of repeated measurements made at high frequencies leading to the Zeno effect. (Right) When an intermediate level is allowed the survival probability of $\ket{A}$, decaying through $\ket{B}$, still retains the Zeno effect. Additionally, the dotted line show the effect of repeated measurements made at lower frequencies leading to the anti-Zeno effect. The vertical axes are on log scale both here and in the previous figure.\label{zeno}}
\end{figure}

Skipping the details, we move to the results, which are shown in Figs.~\ref{zeno}. We plot the survival probability as a function of time of $\ket{A}$ ($P_A$) and compare it to the survival probability of $\ket{B}$ ($P_B$). In the latter case, there is no intermediate level because we cut the coupling to the first level by letting $\Omega\rightarrow0$ as seen from Eq.~(\ref{e1}). The ``unmeasured" survival probability of the states $\ket{A}$ and $\ket{B}$ are the dashed curves and in both cases for long times the decay is exponential. However, the difference between these two cases clearly shows that the presence of a second bound state is crucial for the anti-Zeno effect.

In the Raizen experiment~\cite{PhysRevLett.87.040402}, the system is interrupted by changing the acceleration. In our model, this is achieved by shifting the value of $\Omega$. When $\Omega$ is much smaller than the difference between $E_A$ and $E_B$, there are no oscillations between $\ket{A}$ and $\ket{B}$ and $\ket{A}$ becomes stable. During the interruptions, $\ket{B}$ still decays to the continuum, but the population of $\ket{A}$ remains the constant. Thus, by tuning the parameters of the Hamiltonian we can perform measurements just as in the experiment. Remarkably, our results look almost identical to those of the experiment. This, at least in our minds, clearly shows that anti-Zeno effect has nothing to do with non-projective nature of measurements.

Sudarshan's insight about quantum mechanics was spot on. A model that he had studies four decades before~\cite{bib10} was able to explain away a rather confusing situation in a simple way. He helped Anil and myself every step of the way in finishing this paper. After the first version was completed Anil finished his PhD and left for New Mexico to join Carlton Caves' group as a postdoctoral fellow. During this time Sudarshan helped me to keep the paper moving. (We had very hard reports from PRL after the first submission.) Nevertheless, Sudarshan declined to be an author on the paper. Looking back, I don't understand why, but it's likely that he wanted Anil to take charge and supervise a student on his own. And in the end, while Sudarshan was around to discuss science on this matter, he was rather hands-off. Nevertheless, I look back at this project with a great deal fondness. Even though this paper~\cite{kmzeno} has only been cited less than ten times, I find it to be solid science. A simple model clearly explaining a confusing situation.

\section{Quantum optics and 2005 Nobel Prize}

After the Zeno project was completed (but not published) I began studying the topics of entanglement. We were not a traditional quantum information group and in some ways always took a non-traditional approach. I spent a year being hopelessly lost, though I learned a lot along the way. One reason for studying entanglement was to understand how the dynamics of initially entangled systems can be described. Then we had a pause for a couple of months.

One morning when I got to the office, there was a lot of murmur in the hallways. The Nobel Prize was announced and Glauber was awarded one-half of the prize for his discovery of \emph{quantum optics}. Cesar and I found Sudarshan in his office flat on the sofa. That afternoon he just talked about the history of how the quantum theory of optics was realised. Here's my understanding of this history.

The theory of optics was a mess in the late 1950s and early 1960s. There was the Hanbury Brown and Twiss experiment of intensity correlations. And then there was the discovery of laser. The latter was supposed to be highly quantum, but it wasn't clear which traits of the laser were quantum and which were classical. In fact, it all looked classical. In February of 1963, Glauber published a paper in Physical Review Letters suggesting that coherent states may be a good way to represent the quantum state of light~\cite{PhysRevLett.10.84}. However, Glauber failed to realise that coherent states form an over-complete basis and can diagonally represent any quantum state. In other words, he suggested a non-diagonal representation. This is important because any basis can be used for a non-diagonal representation!

Upon seeing this paper Emil Wolfe approached Sudarshan, who right away saw the solution. He was made to write this paper that very afternoon by Wolf. (Sudarshan complained that he couldn't even have lunch till the paper was finished.) The result was a two and a half page letter that was published in March 1963~\cite{PhysRevLett.10.277}. Here Sudarshan clearly lays out equivalence between classical optics and quantum optics. It goes as the following: represent the state in the coherent state basis. This is done diagonally, as the basis is over-complete. If the corresponding weights are positive than the state of light can be seen as mixtures of coherence sources. However, there are states of light that cannot represent with positive weights. For instance, a Fock state of a definite number of photons. Thus this is a highly quantum state. Sudarshan goes on to lay out the recipe for measuring the quantum nature of light that cannot be described classically.

After Sudarshan's paper, Glauber wrote a long paper in Physical Review~\cite{PhysRev.131.2766}, where he also notes the diagonal representation. There he says 
\begin{quotation}
\emph{During the completion of the present paper note by Sudarshan has appeared which deals with some of the problems of photon statistics that have been treated here. Sudarshan has observed the existence of what we have called the P representation of the density operator and has stated its connection with the representation based on the $n$-quantum states. To that extent, his work agrees with ours in Secs. VII and IX. He has, however, made a number of statements which appear to attach an altogether different interpretation to the P-representation. In particular, he regards its existence as demonstrating the ``complete equivalence" of the classical and quantum mechanical approaches to photon statistics. He states further that there is a ``one-to-one correspondence" between the weight functions P and the probability distributions for the field amplitudes of classical theory.}
\end{quotation}
He repeats this statement in his Nobel Lecture, i.e., that P-function does not exists for many cases, see around minute 40~\cite{nobel}. 

Sudarshan's point was simple. Whenever there the diagonal function (P-function) is positive there is a probabilistic interpretation and the light should be thought to be classical. This is very much like how we think of separable states. On the other hand, there are clear cases where the diagonal function cannot be positive and there is no classical explanation for the state of the light. One wonders if this means that the Nobel Prize committee took Glauber's viewpoint on this matter. Probably not. The big difference, as far as I can tell, is that Sudarshan moved on to other things in physics. As one colleague put it, Sudarshan composed a great album but didn't go around giving concerts to promote the album. He just got on with making the next album. Glauber, on the other hand, kept playing the same old quantum optics tune for the next four decades.

\section{Open Dynamics}
Even before delving into quantum optics Sudarshan had been thinking about quantum stochastic processes. The idea was to describe the dynamics of a system when it interacts with an environment. Consider a bipartite state $\rho^{\mathcal{SE}}$ of the system (labeled by $\mathcal{S}$) and the environment (labeled by $\mathcal{E}$). The total unitary evolution is as follows
\begin{equation}
\rho^{\mathcal{SE}}(t)=U\rho^{\mathcal{SE}}(t_0)U^\dag.
\end{equation}
The dynamics are considered open when we cannot observe the state of the environment and do not know the details of interaction unitary. However, if we had the knowledge of the state of the environment and the unitary transformations, by calculating the closed evolution we would know what the state of the system will be at any point by tracing over the environment
\begin{equation}
\rho^{\mathcal{S}}(\cdot)=\mbox{Tr}_{\mathcal E}[\rho^{\mathcal{SE}}(\cdot)].
\end{equation}
Sudarshan noticed that this not all that different from classical a stochastic process, which is described by the stochastic matrix. He, with two co-authors, began forming the quantum analogue of the classical stochastic matrix in Ref.~\cite{PhysRev.121.920}. This is known as the dynamical map formalism.

\subsection{Dynamical Map Formalism}

Suppose we are not interested in the dynamics of the system for all times but rather in the transformation between two times, i.e., the evolution of the system from time $t_0$ to time $t$. Then we need to define a mapping from density matrices to density matrices such that all allowed initial states of the system are mapped to the corresponding final states in a linear fashion. We closely follow the arguments originally put forth by Sudarshan et al.~\cite{PhysRev.121.920}.

\textbf{The $\mathcal{A}$-form.} Consider an operator acting on the density matrix mapping the state to another density matrix linearly
\begin{equation}
\rho^{\mathcal{S}}_{r's'}(t_0)\rightarrow\mathcal{A}_{rs;r's'}\rho^{\mathcal{S}}_{r's'}(t_0)=\rho^{\mathcal{S}}_{rs}(t).
\end{equation}
Above $\rho^{\mathcal{S}}$ is labeled by two indices and the operator $\mathcal{A}$ is labeled by four, meaning if $\rho^{\mathcal{S}}$ is a $d\times d$ matrix then $\mathcal{A}$ is $d^2\times d^2$ matrix. We can think of the above equation as a super matrix $\mathcal{A}$ acting on a column vector $\rho^{\mathcal{S}}$. This is very similar to a classical stochastic process, where a stochastic matrix maps a probability vector to another probability vector. Matrix is $\mathcal{A}$ called the stochastic map~\cite{PhysRev.121.920, Davies70}.

The only restriction we need to place on the stochastic map is that it map a density matrix to another density matrix. This implies that it must preserve trace, Hermiticity, and positivity of the density matrix. These restriction translate into the following properties for $\mathcal{A}$
\begin{center}
\begin{tabular}{l l}
$\mathcal{A}_{nn,r's'}=\delta_{r's'}$ \hspace{1cm} 
& Trace preservation, \cr
$\mathcal{A}_{rs,r's'}=\left(\mathcal{A}_{sr,s'r'}\right)^*$ \hspace{1cm} &
Hermiticity preservation, \cr
$x^*_r x_s\mathcal{A}_{rs,r's'}y_{r'}y^*_{s'}\geq 0$ \hspace{1cm} &
Positivity. \cr
\end{tabular}
\end{center}

\textbf{The $\mathcal{B}$-form.} Following Sudarshan et al.~\cite{PhysRev.121.920} again, let us rearrange the stochastic map to define a dynamical map
\begin{eqnarray}
\mathcal{B}_{rr';ss'}=\mathcal{A}_{rs;r's'}.
\end{eqnarray}
For a $4\times 4$ map, matrix $\mathcal{A}$ tranforms as follows
\begin{eqnarray}\label{atobform}
\left(
\begin{array}{cccc}
\mathcal{A}_{11} &\mathcal{A}_{12} &\mathcal{A}_{13} &\mathcal{A}_{14} \\
\mathcal{A}_{21} &\mathcal{A}_{22} &\mathcal{A}_{23} &\mathcal{A}_{24} \\
\mathcal{A}_{31} &\mathcal{A}_{32} &\mathcal{A}_{33} &\mathcal{A}_{34} \\
\mathcal{A}_{41} &\mathcal{A}_{42} &\mathcal{A}_{43} &\mathcal{A}_{44} \\
\end{array}
\right)
\rightarrow
\left(
\begin{array}{cccc}
\mathcal{A}_{11} &\mathcal{A}_{12} &\mathcal{A}_{21} &\mathcal{A}_{22} \\
\mathcal{A}_{13} &\mathcal{A}_{14} &\mathcal{A}_{23} &\mathcal{A}_{24} \\
\mathcal{A}_{31} &\mathcal{A}_{32} &\mathcal{A}_{41} &\mathcal{A}_{42} \\
\mathcal{A}_{33} &\mathcal{A}_{34} &\mathcal{A}_{43} &\mathcal{A}_{44} \\
\end{array}
\right).
\end{eqnarray}

The properties of the stochastic map translate in the following manner for the dynamical map~\cite{PhysRev.121.920}.
\begin{center}
\begin{tabular}{l l}
$\mathcal{B}_{nr',ns'}=\delta_{r's'}$ \hspace{1cm} 
& Trace preservation, \cr
$\mathcal{B}_{rr',ss'}=\left(\mathcal{B}_{ss',rr'}\right)^*$ \hspace{1cm} &
Hermiticity preservation, \cr
$x^*_r y_{r'}\mathcal{B}_{rr',ss'}x_{s}y^*_{s'}\geq 0$ \hspace{1cm} &
Positivity. \cr
\end{tabular}
\end{center}
To get complete positivity we can replace the last line as 
\begin{equation}
z_{rr'}\mathcal{B}_{rr',ss'}z_{ss'}^*\geq 0 \quad
\mbox{Positivity.}
\end{equation}
A key tool that I learned from Sudarshan was how to manipulate indices. This allowed me to go to the next step, as I will discuss below.

The advantage of writing the dynamical map $\mathcal{B}$ is that it has nicer properties than the stochastic map $\mathcal{A}$. The dynamical map is Hermitian and its diagonal $d\times d$ block elements have a unit trace. For these nicer properties, we have sacrificed the simplicity of the composition of the stochastic map on the state. The stochastic map acts as a matrix on a state that is a column vector, while the dynamical map acts on the state in the following manner
\begin{eqnarray}
\mathcal{B}_{rr';ss'}\rho^{\mathcal{S}}_{r's'}(t_0)
=\rho^{\mathcal{S}}_{rs}(t).
\end{eqnarray}

While this looks complicated, the dynamical map itself comes with a different simple form for the action. Since $\mathcal{B}$ is Hermitian, we can write it in its eigen-form as $\mathcal{B}=\sum_\alpha \lambda_\alpha C_\alpha \times C_\alpha^\dag$. The equation then becomes
\begin{eqnarray}
\mathcal{B}_{rr';ss'}\rho^{\mathcal{S}}_{r's'}(t_0)
=\sum_\alpha \lambda_\alpha [C_\alpha]_{rr'} \rho_{r's'}(t_0) [C_\alpha^*]_{ss'}.
\end{eqnarray}
If the eigenvalues of $\mathcal{B}$, $\{\lambda_\alpha\}$, are positive then we can redefine $C_\alpha \to \sqrt{\lambda_\alpha} C_\alpha$. This is the well-known Kraus form of the map and it predates Kraus's paper on the topic by a decade~\cite{PhysRev.121.920}! Though, one key difference between the two approaches is that Kraus explicitly mentioned complete positivity, while Sudarshan does not, and yet still gets the right result.

Neither Sudarshan (and colleagues) nor Kraus could have predicted that these ideas would become hugely important for quantum information theory. There are some things in history that are mind-boggling. For instance, Kraus studied under German physicist G\"unther Ludwig. Another student of Ludwig was Arno B\"ohm, who was research associate of Sudarshan back in Syracuse. When Sudarshan moved to Texas B\"ohm also moved and became a professor at UT Austin. (B\"ohm's office was right next to Sudarshan's.) In 1980 Kraus visited B\"ohm at UT Austin for one year and gave series of lectures. In the audience was Bill Wootters, who along with B\"ohm and John Dollard, published the corresponding lecture notes~\cite{krausbook}. The prominence of Wootters in the quantum information community ensured that it was Kraus's lecture notes that introduced the ideas of quantum maps to the quantum information community. Ironically, in this instance, Sudarshan lost credit in his own house.

\subsection{Memoryless master equation}

Sudarshan was never the one to stand still. After dynamical maps, Sudarshan kept working on open systems problems but began searching for mechanism for Markovianity. It took another 15 years before an exact formulation of the memoryless master equation was formed~\cite{sudarshangorini, Lindblad1975}. This topic too has rich history, but it's better covered elsewhere~\cite{gksl} and I will move on. However, here too Sudarshan and colleague lost credit to Lindblad. As far as I can tell, Lindblad didn't go around advertising his work, but still somehow came out on the top. In this instance, it's just misfortune.

\section{Non-Markovian dynamics}

One topic that Sudarshan talked about a lot was non-Markovianity, i.e., the converse of the GKSL master equation~\cite{cesarnonmarkov}. In the quantum case, it remained unclear in the 2000s how to describe non-Markovian dynamics. One key issue that needed to be solved was the matter of dynamics in the presence of initial correlations. This is because the initial correlation is a record of past interactions between the system and the environment. If the future state of the system depends on the present correlation, then it also depends on the past~\cite{mazzola2012dynamical, rodriguez2012unification}.

The initial correlation problem was being addressed with full force by the group in Texas when I began working on open dynamics. Based on simple arguments due to Pechukas~\cite{pechukas94a}, they had written several papers arguing against completely positive dynamics~\cite{jordan:052110, jordan06a, rodr, shaji_whos_2005} when the initial system-environment state was entangled. Later it was shown that the problem did not go away for separable states \cite{CarteretTernoZyczkowski05, Rodriguez07a, PhysRevA.81.012313, modi_positivity_2012}.

I was asked to look at quantum process tomography in this context. This is because several experiments were reporting not-completely positive maps~\cite{Howard06, myrskog:013615, PhysRevLett.90.193601, Wein:121.13}. Quantum process tomography is performed by preparing different initial states and sending them through a quantum process. The subsequent output state is measured, and the correlations between preparation and measurement allow for reconstructing the map representing the process.

\subsection{Preparations in quantum mechanics}
\label{prepinqm}

I soon realised that the missing piece of the puzzle was the preparation of the initial state. For a map to be experimentally accessible, one has to do quantum process tomography. The first step in this procedure is to prepare the system in different initial states. However, if the system is initially correlated with the environment, then we cannot prepare it in different initial states. Moreover, there is no unique way to prepare a given state~\cite{modi_preparation_2011}.

In quantum process tomography, a linearly independent set of states that span the system space have to be prepared. But a generic experiment requires preparing a set that is larger than a linearly independent set. As such, a quantum experiment can be described in three general steps. The experiment begins with an unknown (or a fiducial) state that has to be altered into known input state. After the preparation, the prepared state is subjected to some quantum operation, and finally, the outcome is analysed. Preparation procedures are very complicated in practice. Since we cannot describe each preparation procedure in detail, the ideas were to develop a general theory of preparation. 

We can think of preparation as the following: considering an experiment that starts with a generic state of the system $\rho$, which is then altered into a set of desired inputs $P^{(m)}$. We have to somehow connect the initial state to the input states. The most general dynamics of a quantum state are described by a stochastic map. We can denote the procedure for preparing the $m$th state with a map $\mathcal{P}^{(m)}$. The only restriction we put on the preparation map is that it be completely positive. This is because no matter what the initial state of the system is, we should be able to prepare the state. Then a preparation procedure $\mathcal{P}$ can be implemented with the aid of an ancillary system and its action is defined as
\begin{equation}
\mathcal{P}^{(m)}\rho =P^{(m)}.
\end{equation}

It turns out that not all preparation procedures are trace preserving. Consider a polariser plate to prepare a photon in, say, vertical polarisation state. This procedure cannot always succeed. The action of such maps is written as
\begin{equation}
\mathcal{P}^{(m)}\rho=r^{(m)} P^{(m)},
\end{equation}
where the normalising factor 
\begin{equation}
r^{(m)}=\mbox{Tr} [\mathcal{P}^{(m)}\rho]
\end{equation}
is the probability with which $\rho$ will become $P^{(m)}$. However, there is a complete map, $\sum_m \mathcal{P}^{(m)}$, that preserves the trace. This suggests that to prepare $m$ input states, we will, in general, need $m$ preparation maps. In other words, an experiment will require $m$ preparation procedures. Not all of these procedures will be completely unrelated, in fact, most will share some common features.

\textbf{Preparations in open quantum mechanics.} How does the situation above change when we consider open quantum systems? We simply follow the procedure laid out above, but instead of using the generic initial state $\rho$, we must use a bipartite state of the system and the environment $\rho^\mathcal{SE}$. The preparation map still only acts on the state of the system alone:
\begin{equation}
r^{(m)} R^{\mathcal{SE}^{(m)}} = \mathcal{P}^{(m)} \otimes \mathcal{I} (\rho^{\mathcal{SE}}),
\end{equation}
where $R^{\mathcal{SE}^{(m)}}$ is the bipartite state of the system and environment after the preparation. For completeness, in the equation above we have included an identity map $\mathcal{I}$ acting on the state of the environment, but for simplicity, we will omit writing it from here on.

Even though the preparation procedure only acts on the state of the system, the state of the environment, in general, will indirectly be affected by the procedure. Suppose our goal is to prepare an uncorrelated state. Simplifying the notation above we get
\begin{equation}
R^{\mathcal{SE},(m)} =\mathcal{P}^{(m)} (\rho^{\mathcal{SE}}) =P^{(m)}\otimes\rho^{\mathcal{E},(m)}.
\end{equation}
In the last equation, we see that the state of the environment has picked a superscript $m$, because if $\rho^{\mathcal{SE}}$ is initially correlated, then an action on the system part will necessarily affect the state of the environment. However, if the weak coupling assumption is retained, then the preparation procedures for the closed and the open cases are the same. Let us investigate this through two specific procedures. See~\cite{Kuah02} for a similar analysis for quantum measurements.

Note that, if the system evolves in a closed form then all preparation procedures are equivalent. Trouble only arises when the preparation procedure indirectly affects the state of the environment which then interacts with the system. Additionally, even though we will only use the results here in analysing quantum process tomography procedures, they apply to any quantum experiment that interacts with an environment. With these ideas, Sudarshan and I wrote a paper showing that when one accounts for preparation of initial state in a quantum process tomography experiment one can come across not-completely positive maps~\cite{modi_role_2010} due to inconsistencies in preparation procedures when there are initial correlations present.

\subsection{Superchannel}

As we saw above that the additional step of preparation leads to complications for quantum process tomography experiments. At the same time, this step cannot be avoided for systems initially correlated with the environment. A way around was proposed by Aik-Meng Kuah, another PhD student in the group~\cite{kuah:042113}. He found a complicated, yet elegant, process tomography procedure for a single qubit initially with its environment. The caveat was that the initial preparation had to be projective. 

Soon after, I realised that there is a more general way describe dynamics due to initial correlations~\cite{modidis, modi_operational_2012}. Let us try to attack this problem directly by writing down the equation for the process, including the preparation:
\begin{equation}
Q^{(m)}=\frac{1}{r^{(m)}}
\mbox{Tr}_{\mathcal E}[U\mathcal{P}^{(m)}\rho^\mathcal{SE} U^\dag].
\end{equation}
In terms of matrix indices we have
\begin{equation}
Q^{(m)}_{rs}=\frac{1}{r^{(m)}}\sum_{\epsilon}
U_{r\epsilon;r'\alpha}\mathcal{P}^{(m)}_{r'r'';s's''}\rho^\mathcal{SE}_{r''\alpha;s''\beta} U^*_{s\epsilon;s'\beta},
\end{equation}
where the sum over $\epsilon$ denotes the trace with respect to the environment. We are interested in the reduced dynamics of the system states; not so much in the details of the preparations. Since the preparation map only acts on the system, the trace with the environment has no effect on it. Thus, we can just pull the preparation map out of the trace,
\begin{eqnarray}
Q^{(m)}_{rs}
&=&\frac{1}{r^{(m)}}\nonumber
\mathcal{P}^{(m)}_{r'r'';s's''}\sum_{\epsilon}
U_{r\epsilon;r'\alpha}
\rho^\mathcal{SE}_{r''\alpha;s''\beta} 
U^*_{s\epsilon;s'\beta}\\
&=&\frac{1}{r^{(m)}}\label{RawProcessEquation}
\mathcal{M}^{(rs)}_{r'r'';s''s'}
\mathcal{P}^{(m)}_{r'r'';s's''}.
\end{eqnarray}
In the last equation, the matrix $\mathcal{M}$ is defined as:
\begin{equation}\label{mmap}
\mathcal{M}^{(rs)}_{r'r''; s''s'} = \sum_{\epsilon} U_{r \epsilon,r' \alpha} 
{\rho^{\mathcal{SE}}}_{r''\alpha,s''\beta} U^*_{s \epsilon,s' \beta} .
\end{equation}
Note that in Eq.~\ref{RawProcessEquation} the superscript indices on $\mathcal{M}$ match the elements on the left hand side of the equation, while the subscript indices are summed on the right hand side of the equation. 

The output state, $Q^{(m)}$, is given by the matrix $\mathcal{M}$ acting generally on the preparation of the input state, $\mathcal{P}^{(m)}$. Therefore, matrix $\mathcal{M}$ fully describes the process before any preparation is made. $\mathcal{M}$ contains both $U$ and $\rho^{\mathcal{SE}}$; however knowing $\mathcal{M}$ is not sufficient to determine $U$ and $\rho^{\mathcal{SE}}$. As expected, it should not be possible to determine $U$ and $\rho^{\mathcal{SE}}$ through measurements and preparations on the system alone without access to the environment. Conversely, $\mathcal{M}$ contains all information necessary to fully determine the output state for any prepared state. 

I was then able to prove some nice properties for $\mathcal{M}$. Namely, trace preservation, Hermiticity, and complete positivity of $\mathcal{M}$-map. Indeed, this shows that the dynamics of initially correlated systems are also completely positive! I couldn't have done this without learning to manipulate indices, a trick I learnt from Sudarshan. This construction is operationally meaningful, which is a plus on the attempts to use not-completely positive maps to describe such dynamics. It's worth mentioning that the way this method gets around the theorem due to Pechukas~\cite{pechukas94a} is it no longer treats the initial state input of the problem. This is sensible because if want to vary the initial state we have to apply a preparation procedure. Then why not simply treat the preparation procedure as the input of the map! Interestingly, Alicki, in his response to Pechukas almost had this construction~\cite{PhysRevLett.75.3020}.

\subsection{Process Tensor and Higher Order Maps}

In my PhD thesis map $\mathcal{M}$ was called the \emph{dynamical $\mathcal{M}$-map} or just \emph{$\mathcal{M}$-map} for short. Later, it was experimentally tested and reported in Ref.~\cite{PhysRevLett.114.090402}, where it was dubbed as the \emph{superchannel}. Superchannel is a type of higher-order map and a close cousin of the supermap~\cite{chiribella_transforming_2008}, which was invented around the same time as the superchannel. I did not know this at the time. There were other inventions~\cite{chiribella_quantum_2008, chiribella_theoretical_2009} that I also was not aware of, but they too became critical for describing non-Markovian quantum processes.

That is, generalising the superchannel to multi-time steps naturally leads to a complete theory of non-Markovian quantum processes~\cite{PhysRevA.97.012127, 1367-2630-18-6-063032}, which is called the \emph{process tensor} framework. We've learnt since then that these ideas were explored long ago by Lindblad~\cite{lindblad_non-markovian_1979}, Accardi et al.~\cite{accardi_quantum_1982}, and by Kretschmann and Werner~\cite{kretschmann_quantum_2005} more recently, see here~\cite{MilzReview} for a mini-review. 

Nevertheless, this way of looking at quantum processes naturally resolves the ambiguity of what makes a quantum processes Markovian~\cite{PhysRevLett.122.160401} and when the memory is quantum~\cite{arXiv:1811.03722}. It leads to a unifying framework for spatio-temporal correlation~\cite{arXiv:1710.01776, kolmogorov}, where a space-time version of the Born rule appears~\cite{arXiv:1702.01845}. Later we also generalised Kuah's idea~\cite{kuah:042113} to fit restricted control process tensor~\cite{PhysRevA.98.012108}. These ideas allow us to study the structure of memory in quantum processes~\cite{PhysRevLett.120.040405, PhysRevLett.122.140401, PhysRevA.99.042108, simon-div}, and quantitatively relate it the memory-kernel master equations~\cite{Pollock2018tomographically}. Lastly, we are now able to go back to the foundational question that Sudarshan cared about: why is nature so Markovian~\cite{FigueroaRomero2019almostmarkovian}.

\section{Conclusions}

I learnt a lot in my years as a PhD student under George Sudarshan. In recent years we had not kept much contact. Last time I met him was in Torun for the Symposium honouring 40 years of Gorini-Kossakowski-Sudarshan-Lindblad master equation. He gave a great talk there remembering the old times. However, it was tough to talk to him about research. I would have really like to hear his views on the process tensor formalism and the many things we've done with it in the last few years. I remember his remark after reading my thesis chapter on the superchannel. He found a great deal of clarity in it. Though at the time I didn't know what to do with it. It took another five years before I realise the potential it carried. There is still a lot to do, and I feel that I got an extra boost by learning from Sudarshan.

\bibliographystyle{unsrt}
\bibliography{ECG.bib}

\end{document}